# Positron clouds within thunderstorms

Joseph R. Dwyer[1]†, David M. Smith[2], Bryna J. Hazelton[3], Brian W. Grefenstette[4], Nicole A. Kelley[5], Alexander W. Lowell[5], Meagan M. Schaal[6] and Hamid K. Rassoul[7]

[1]Department of Physics and Space Science Center (EOS), The University of New Hampshire, Durham, NH 03824, USA

[2]Physics Department and Santa Cruz Institute for Particle Physics, University of California, Santa Cruz, CA 95064, USA

[3]Department of Physics, University of Washington, Seattle, Washington, USA 98195.

[4]Space Radiation Laboratory, California Institute of Technology, Pasadena, California, USA 91125.

[5]Space Sciences Laboratory, University of California, Berkeley, California, USA 94720.

[6]National Academies of Science, resident at the Naval Research Laboratory, Washington, DC 20375

[7]Department of Physics and Space Sciences, Florida Institute of Technology, Melbourne, FL 32901, USA

We report the observation of two isolated clouds of positrons inside an active thunderstorm. These observations were made by the Airborne Detector for Energetic Lightning Emissions (ADELE), an array of six gamma-ray detectors, which flew on a Gulfstream V jet aircraft through the top of an active thunderstorm in August 2009. ADELE recorded two 511 keV gamma-ray count rate enhancements, 35 seconds apart, each lasting approximately 0.2 seconds. The enhancements, which were about a factor of 12 above background, were both accompanied by electrical activity as measured by a flat-plate antenna on the underside of the aircraft. The energy spectra were consistent with a source mostly composed of positron annihilation gamma rays, with a prominent 511 keV line clearly visible in the data. Model fits to the data suggest that the aircraft was briefly immersed in clouds of positrons, more than a kilometer across. It is not clear how the positron clouds were created within the thunderstorm, but it is possible they were caused by the presence of the aircraft in the electrified environment.

† Email address for correspondence: Joseph.Dwyer@unh.edu

# 1. Introduction and previous observations

The physics of particle transport in gaseous media with strong electric fields has proven to be a rich topic with many surprising new phenomena discovered in recent years. It has been established that electric fields generated by both thunderstorms and lightning accelerate electrons to relativistic energies, emitting x-rays and gamma rays. Specifically, lightning leaders have been observed to emit bright sub-microsecond pulses of x-rays with energies typically in the few hundred keV range (Moore et al. 2001; Dwyer et al. 2003, 2004a, 2005a, 2011; Yoshida et al. 2008; Howard et al. 2008, 2010; Chubenko et al. 2009; Hill et al. 2012; Hill 2012; Mallick et al. 2012; Saleh et al. 2008; Schaal et al. 2012, 2013, 2014). Long laboratory sparks have been shown to produce similar x-ray emissions (Dwyer et al. 2005b, 2008a; Kostyrya et al. 2006; Rahman et al. 2008; Rep'ev and Repin 2008; Nguyen et al. 2008, 2010; Nguyen 2012; Babich and Loĭko 2009; March and Montanyà 2010, 2011; March et al. 2012; Gurevich et al. 2011). Thunderstorms produce bright sub-millisecond bursts of gamma rays, called terrestrial gamma-ray flashes (TGFs), with energies reaching several tens of MeV (Fishman et al. 1994, 2011; Smith et al. 2005; Cummer et al. 2005, 2011; Dwyer and Smith 2005; Grefenstette et al. 2008; Briggs et al. 2010; Connaughton, et al. 2010, 2013; Gjesteland et al. 2010, 2011, 2012; Marisaldi et al. 2010a,b; Tavani et al. 2011; Østgaard et al. 2012). Terrestrial gamma-ray flashes are usually observed by spacecraft in low-Earth orbit, but have also been observed by an aircraft and on the ground (Dwyer et al. 2004b; Smith et al. 2011b; Dwyer et al. 2012). They are so bright that they may be radiation hazards to individuals in aircraft (Dwyer et al. 2010). For both the lightning/laboratory spark emissions and TGFs, the x-rays and gamma rays are produced by bremsstrahlung interactions of energetic electrons with air. However, it is a theoretical challenge to explain how so many high-energy electrons are generated in our atmosphere on such short time scales.

As the TGF gamma rays propagate up and out of the atmosphere, they also generate energetic secondary electrons via photoelectric absorption and Compton scattering and energetic secondary electrons and positrons via pair production. Some of these electrons and positrons escape to space, travelling along the geomagnetic field line where they may be observed by spacecraft. These events are called terrestrial electron beams (TEBs) and have been observed by the CGRO, RHESSI and Fermi spacecraft (Dwyer et al. 2008b; Cohen et al. 2010; Briggs et al. 2011; Carlson et al. 2011).

Another kind of emission from thunderclouds is the gamma-ray glow. Gamma-ray glows appear as sub-second to minute long emissions of gamma rays (Torii et al. 2004; Tsuchiya et al., 2007; Chilingarian et al. 2010; Babich et al. 2010). Like TGFs, the glows are produced by bremsstrahlung emissions from energetic electrons and in some cases have been found to have spectra similar to those of TGFs. However, glows last much longer than TGFs and have much lower fluxes. Gamma-ray glows have been observed by aircraft, balloon and on the ground. For example, using a NASA F-106 jet carrying NaI scintillation detectors, Parks et al. (1981) and McCarthy and Parks (1985) demonstrated that active thunderstorms produce gamma rays (originally referred to as x-rays) that last tens of seconds, with energies greater than 110 keV. They found that the gamma-ray emissions were generally terminated, rather than caused, by lightning. In a series of balloon flights, Eack et al. (1996a, 1996b, 2000) flew scintillators and electric field sensors through and above active thunderstorms and measured gamma-ray glows of up to 120 keV in energy. They found that the gamma-ray emissions occurred at an altitude of 4 km where the electric field was highest (Eack et al. 1996a). The emission persisted while the balloon passed through the strong-electric-field region within the storm, except that it terminated and then restarted following two lightning flashes. In other balloon soundings by Eack et al., similar high gamma-ray fluxes were recorded in an anvil at 14 km and above the thundercloud at 15 km altitude (Eack et al. 2000,1996b).

Gamma-ray glows have been measured in regions of Japan from thunderclouds with low charge centers (e.g., Torii et al. 2002, 2011; Tsuchiyu et al. 2007, 2011) and on high mountains (Brunetti et al. 2000; Chubenko et al. 2000, 2003; Alexeenko et al. 2002; Torii et al. 2009; Tsuchiya et al. 2009; Chilingarian et al. 2010, 2012a, 2012b, 2013). For example, Torii et al. (2002) measured gamma-ray enhancements of up to 70 times the local background level at the Monju nuclear reactor in Japan during a winter thunderstorm. Chilingarian et al. (2010) reported that energetic particle count rates at Mount Aragats more than doubled when a thundercloud was 100-200 m above the detector. For more details about earlier observations, we refer the reader to Suszcynsky et al. (1996), Dwyer, Smith and Cummer (2012) and Dwyer and Uman (2014).

## 2. Runaway electrons

In this paper we present two cases of an aircraft suddenly being immersed in positron-annihilation radiation. Each potential source of the positrons that will be considered depends on the same fundamental physics, which we outline in this section.

The key building block for producing energetic radiation from thunderstorms and lightning is the runaway electron, first identified by Wilson in 1925 (Wilson 1925). When an electric field is present in air, or any gaseous medium, electrons will run away when the rate of energy gain from the electric field exceeds that rate of energy loss from interactions with the gas (see figure 1). For energetic electrons, ionization and bremsstrahlung emission are the dominant sources of energy loss.

As can be seen in figure 1, an electron (or positron) may run away if it has an initial kinetic energy above the threshold, $\varepsilon_{th}$. Such seed particles may be supplied by atmospheric cosmic rays or radioactive decays. In addition, as the runaway electrons propagate, they will experience hard elastic scatters with atomic electrons (Møller scattering), resulting in energetic secondary electrons ejected from the atoms. If the energies of these secondary electrons are above $\varepsilon_{th}$, they may also run away. In this way, more and more runaway electrons are created, resulting in a Relativistic Runaway Electron Avalanche (RREA) (Gurevich et al. 1992; Roussel-Dupré and Gurevich 1996; Lehtinen et al. 1999; Gurevich and Zybin 2001; Dwyer 2003; Babich et al. 2004;

Roussel-Dupré et al. 2008; Milikh and Roussel-Dupré 2010). Simulations have found the threshold electric field for RREAs to develop is $E_{th} = 2.84 \times 10^5$ V/m $\times n$, where $n$ is the density of air relative to that at sea level (Dwyer 2003; Babich et al. 2004), a field value within the range of measured thunderstorm electric fields (McGorman and Rust 1998; Rakov and Uman 2003). Because of elastic scattering of the electrons around the electric field line, $E_{th}$ is slightly larger than the minimum of the energy loss curve ($eE_b$) seen in figure 1.

For very strong electric fields, with $E > E_c$, because the energy loss rate is insufficient to prevent electrons from accelerating, nearly all free electrons will run away, potentially reaching relativistic energies. This is the so-called cold runaway electron mechanism (also called thermal runaway or high-field runaway) and may explain the x-ray emissions from lightning and laboratory sparks (Gurevich 1961; Dwyer 2004; Moss et al. 2006).

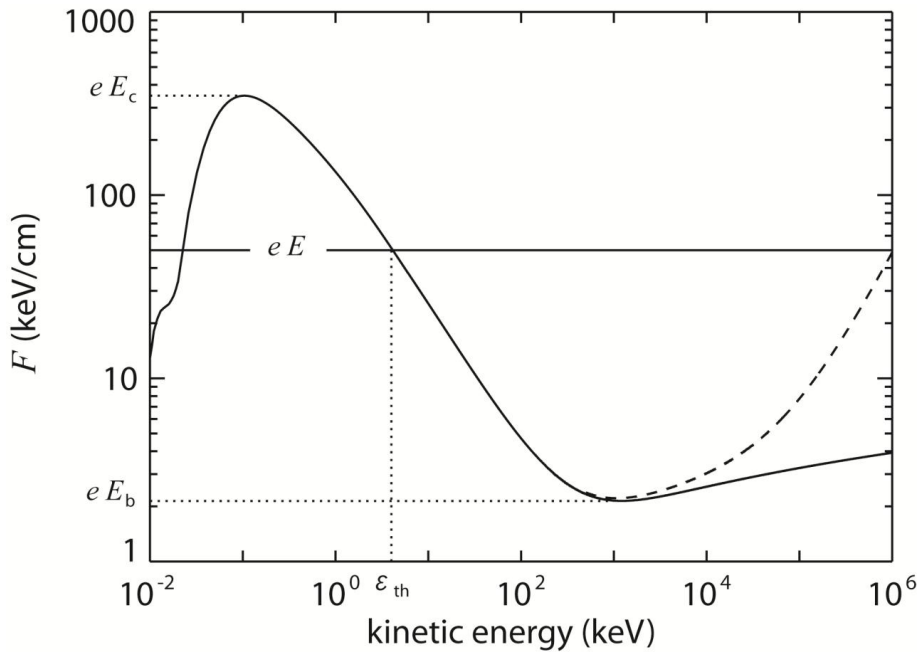

FIGURE 1. Energy loss per unit length of an energetic electron moving through air at standard conditions. The solid curve shows the energy losses due to ionization and atomic excitations. The dashed curve shows the losses due to bremsstrahlung emissions. The solid horizontal line shows an example of the energy gain per unit length from an external electric field. When the rate of energy gain from an electric field is greater than the rate of energy loss, the electrons will "run away." From Dwyer (2004).

Finally, when the propagation of the x-rays, gamma rays and runaway positrons is considered, it has been shown that a positive feedback effect may result, with backward propagating positrons and backscattered x-rays and gamma rays seeding new runaway electron avalanches. This mechanism, called relativistic feedback, could explain the large number of energetic particles generated during TGFs (Dwyer 2003; 2007; 2012; Babich et al. 2005; Liu and Dwyer 2013; Skeltved et al. 2014). Relativistic feedback effects could also be important for gamma-ray glows, especially for the brightest glows. For relativistic feedback, an understanding of the propagation of the positrons is very important, since in most cases it is the positrons that cause the feedback. It has been suggested that runaway positrons can not only generate the explosive growth of energetic particles, but also may help to regulate the electric field inside thunderstorms (Dwyer 2003).

Bright gamma-ray glows are probably produced by electrified regions of thunderstorms in which a substantial amount of RREA multiplication is occurring. For smaller-electric-field regions, the RREA will mostly be seeded by atmospheric cosmic-ray particles. In this case the thunderstorm is acting like gigantic gas proportional counter, enhancing the number of energetic electrons by a factor that could reach thousands. As a thunderstorm charges, more and more of the RREAs will be generated by relativistic feedback (mostly positron feedback), increasing the intensity of the gamma ray glow. As the intensity increases, the discharge current will increase from the elevated conductivity caused by ionization. This discharge may limit the intensity of the glow and eventually cause it to cease. On the other hand, for cases where the thunderstorm electric field is too low to produce runaway electrons and positrons, i.e., $E < E_{\text{th}}$, the electric field can still modify cosmic-ray air showers and the cosmic-ray background, altering the momenta and energies of the charged particles. This effect could be responsible for some enhancements seen on the ground. For example, Chilingarian et al. (2013) observed gamma-ray enhancements on a mountain (which they called thunderstorm ground enhancements, TGEs) and differentiated between events in which runaway electrons are produced and events in which the spectrum of atmospheric cosmic rays is modified by the thunderstorm electric field, referred to as modification of the energy spectra (MOS).

## 3. Airborne Detector for Energetic Lightning Emissions

To study the production of energetic radiation by thunderstorms, the Airborne Detector for Energetic Lightning Emissions (ADELE) was flown on an NCAR/NSF Gulfstream V (GV) jet aircraft in and around Florida in August and September of 2009. ADELE consisted of two sensor heads (referred to here as top and bottom), each containing a 12.7 cm diameter by 12.7 cm long cylinder of plastic (BC-408) scintillator attached to a photomultiplier tube (PMT) detector, a 12.7 cm diameter by 12.7 cm long cylinder of NaI(Tl) attached to a PMT detector and a small 2.54 cm by 2.54 cm plastic scintillation/PMT detector for measuring high fluxes. To help discriminate between upward and downward moving particles (photons, electrons and positrons), lead attenuator sheets (0.3 cm thick) were wrapped around the bottom halves of all the detectors in the top sensor head and around the top halves of all the detectors in the bottom sensor head.

The NaI/PMTs provided good stopping power and energy resolution (e.g., <10% FWHM at 662 keV). They were read out by direct flash digitization of the PMT signal with a Gage Octopus TM PCI card, sampling at 12.5 MHz. These data were continuously recorded in a circular buffer on the card, but were only stored during 1 sec intervals when a trigger was generated by high count rates in the plastic detectors. A fast flat-plate antenna consisted of a 30 cm by 30 cm aluminum plate, mounted on the bottom of the aircraft and terminated with 50 Ohms. Signals from the flat-plate antenna were digitized simultaneously in another channel of the Gage card and recorded along with the two NaI/PMT channels. The antenna measured the derivative of the electric field (*dE/dt*), with a full range of ±12 kV/(m μsec), making it sensitive primarily to local lightning activity.

Because the NaI/PMT detectors saturate at count rates above $\sim10^5$ counts/s, the plastic scintillators were designed to operate at higher rates (up to $3\times10^6$ counts/s). Signals from these detectors were sent to a network of clamping amplifiers with no shaping, and the amplified signals were then split out to four discriminators with threshold energies set to ~50 keV, 300 keV, 1 MeV and 5 MeV. The discriminator rates from the plastic detectors were recorded continuously and, unlike the NaI detectors, did not require a trigger for the data to be saved.

## 4. Thunderstorm observations by ADELE

During the August and September of 2009 campaign, which consisted of 9 flights lasting 37 hours in Colorado and Florida, ADELE recorded at least 12 gamma-ray glows and one TGF (Smith et al. 2011a, 2011b; Dwyer et al. 2012). In particular, on 21 August 2009, the GV was flying at an altitude of 14.1 km over coastal estuaries on the southeast coast of Georgia near the town of Brunswick when it inadvertently entered the upper part of an active thunderstorm cell. The plane experienced moderate to severe turbulence with rapid altitude changes, but no lightning strikes to the plane were noticed. While inside the thunderstorm, ADELE recorded four gamma-ray events, which are shown in figure 2. The last event, labeled 4 in figure 2, was the brightest gamma-ray glow recorded by ADELE. During this time period, ADELE triggered twice, saving the data from the NaI/PMT detectors and the flat-plate antenna. The triggered events are labeled 1 and 3 in the figure. While the data were being saved, ADELE could not record data from additional triggers. As a result, NaI/PMT and flat-plate antenna data were not recorded for the events labeled 2 and 4. For these two events, only crude energy spectra were available from the plastic scintillation detectors, which are presented next.

Figure 2 shows the background subtracted count rates recorded by the plastic scintillator for two energy channels, 300 − 1000 keV (red data) and >1000 keV (blue data). Constant background rates of 1000 counts/sec and 1700 counts/sec were subtracted from the 300 − 1000 keV and >1000 keV data, respectively. As can be seen, the first three events appear to be very similar to each other, with matching time profiles and larger enhancements in the low-energy channel compared with the high-energy channel. These are in contrast to event 4, which is much brighter, lasts longer and has similar enhancements in both the low and high-energy bands. An analysis of event 4 suggests that ADELE entered a downward beam of runaway electrons, i.e., the source region of a gamma-ray glow. As will be shown below, for events 1 and 3, the gamma-ray enhancements in the 300- 1000 keV channel were almost entirely due to 511 keV emissions. Although the energy spectrum for event 2 is not available because the instrument did not trigger, based upon the similarities of the count rate profiles and the similar channel ratios for events 1, 2 and 3, we conclude that event 2 was likely to also be a 511 keV enhancement similar to events 1 and 3.

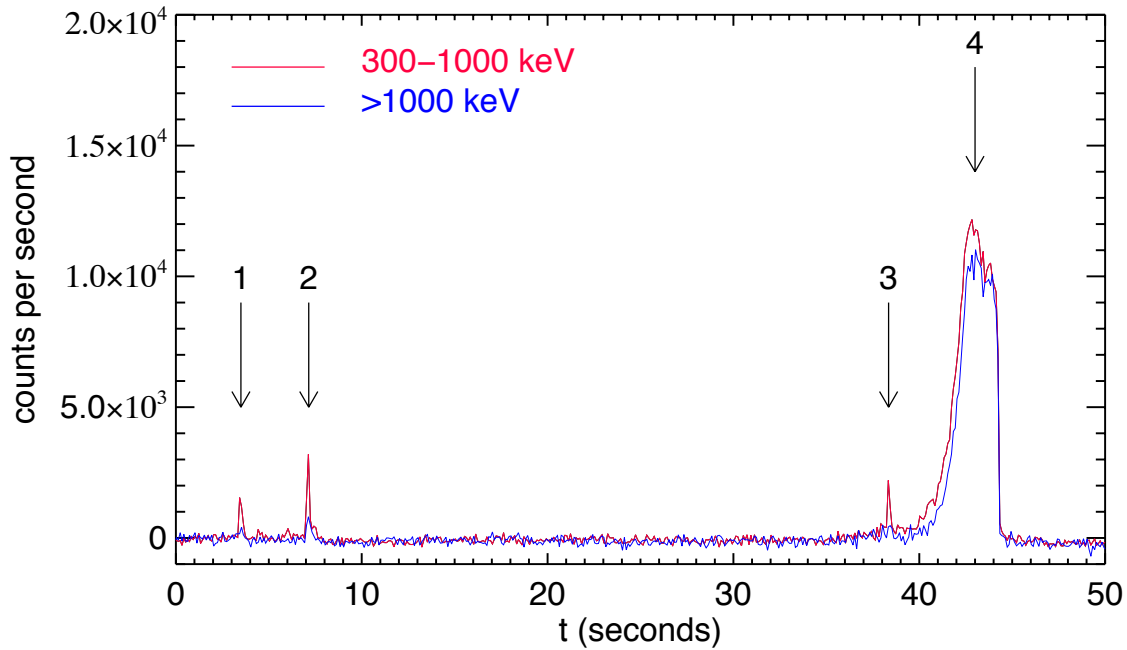

FIGURE 2. Background subtracted count rates in two energy channels recorded by the ADELE plastic scintillation detector versus time when ADELE was inside the thunderstorm on 21 August 2009. Four gamma ray events are labeled 1 – 4. The ratios of the low-energy channel to the high-energy channel were quite different for the first 3 events compared to the last event, showing that for these 3 events the emission is dominated by lower energy photons. Events 1 and 3 are the ones with triggered data and so are the subject of this paper.

Figure 3 shows the 1 second of the NaI/PMT detector data (bottom panel) and the flat-plate antenna data (top panel) for the two triggered events (events 1 and 3). The trigger time for both events is at 0.8 seconds in the records, which is why both count rate enhancements occur at that same time. It should be noted that the two events, which are separated by 35 seconds and approximately a 7.8 km distance, appear surprisingly similar in terms of the count rate time profiles, peak rates and the electrical discharge activity seen with the flat-plate antenna. To within statistical uncertainties, both gamma-ray enhancements appear to begin at the same time as the large deflections in the flat-plate antenna data at 0.787 sec in the records. If we integrate the electric field derivatives from the flat-plate antenna, unphysical large electric fields are inferred, suggesting that the antenna was experiencing electrical discharges during the events.

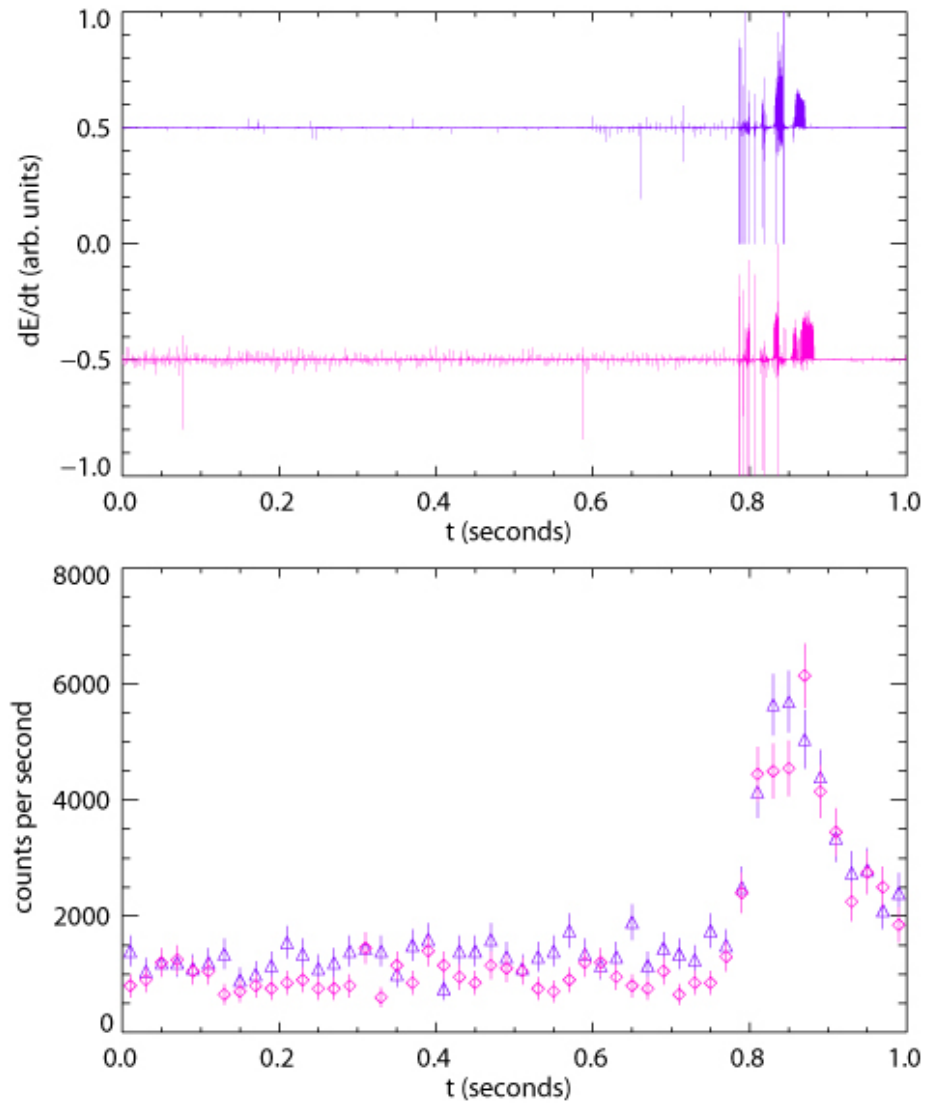

FIGURE 3. *Top panel:* The electric field derivative (*dE/dt*) recorded by a flat-plate antenna on the underside of the GV aircraft for events 1 and 3 shown in figure 2. Event 1 is the lower pink waveform and event 3 is the upper purple waveform. Note the two waveforms have been shifted vertically so that they may be seen more clearly. *Bottom panel:* The $400 - 600$ keV gamma ray count rates from the top and bottom NaI detectors for the two events. The pink diamonds are event 1, and the purple triangles are event 3.

A scatter plot of the deposited energies versus time for the two events is shown in figure 4. The data points are the deposited energies of individual photons (or particles) recorded by the top NaI detector. As can be seen, a very clear and prominent excess of counts occurs at and around 511 keV (yellow dashed line), the positron two-photon annihilation line. There is only a slight enhancement at energies above the annihilation line, but there is large enhancement at lower energies. As can be seen in the figure, the energy spectrum of the background appears to remain constant with time prior to the start of the event at approximately 0.8 sec, and then at 0.8 sec the enhancement appears simultaneously at both 511 keV and at low energies. However, as we shall show below, the low-energy enhancement is consistent with Compton scattering of the 511 keV photons in the material in and around the detectors (i.e., in the aircraft and the air surrounding it). An analysis of the events separately shows that they have nearly identical spectra.

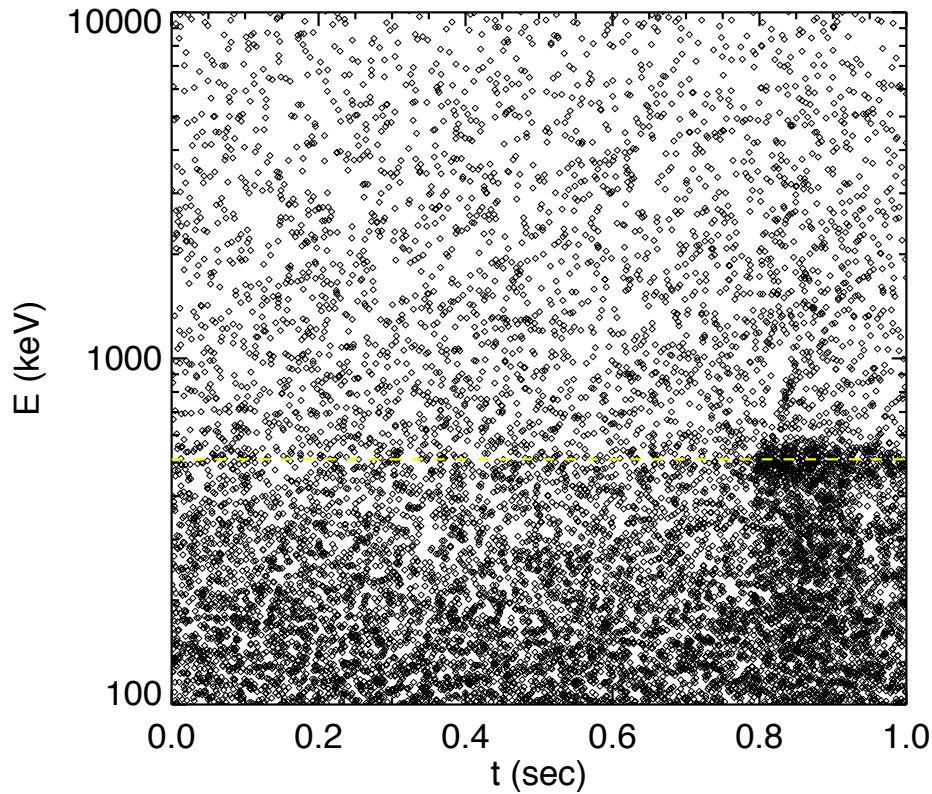

FIGURE 4. Scatter plot of detected photon energies versus time for the two events combined. The data are from the top NaI detector. The yellow dashed line shows 511 keV.

Figure 5 is the energy spectra of the two events combined. The data include both the top and bottom NaI detectors. The blue triangles show the energy spectrum for the times $0 - 0.75$ sec in the records, before the gamma-ray enhancements. We take this spectrum to represent the local background at the time of the events. The black diamonds are the data for times $0.8 - 0.9$ sec, during the peak of the 511 keV enhancements. The spectra were divided by the time periods over which the data were accumulated to give the counts per unit energy per second.

To estimate how much the 511 keV line was enhanced above the background value, the continuum below the 511 keV peak must be subtracted.  This was done for each spectrum by fitting a power-law to the continuum at energies above 700 keV and then subtracting this power law component below the 511 keV peaks.   Using this method, we found that the count rate of 511 keV photons was 12±2 times larger than the local background during the enhancements.  In contrast, the count rate enhancement above 700 keV, away from the 511 keV line, was only a factor of 1.52±0.03 above background.   As will be discussed next, the enhancements at energies below 511 keV are consistent with Compton scattered 511 keV photons and so may be considered part of the positron annihilation emission.

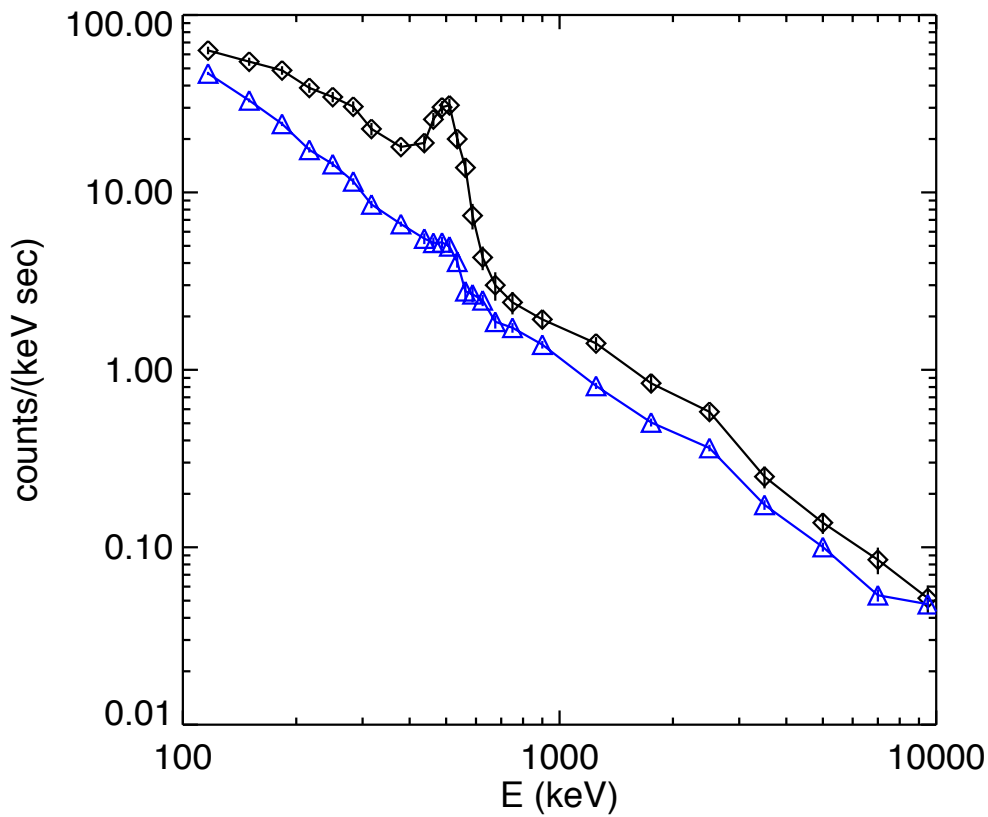

FIGURE 5. Combined energy spectra of the two events during times 0.8 − 0.9 sec (black diamonds) and times 0 −0.75 sec (blue triangles) in the records.  The data are from both the top and bottom NaI detectors on ADELE.

Figure 6 shows the background-subtracted spectrum for the two events combined, found by subtracting the background data seen in figure 5 (blue triangles) from the data for the 511 keV enhancement (black diamonds).  The data in figure 6 are for times 0.8 − 0.9 sec and are from both the top and bottom NaI detectors.  When looking at the top and bottom detectors separately, the background-subtracted spectra are very similar, with a top to bottom count rate ratio of

0.86±0.09 in the 200 – 400 energy range and 1.17±0.11 in the 400 – 600 keV range. The ratio for 100 – 600 keV is 0.99±0.07, showing that to within statistical uncertainties the counts arriving from below and above are the same. The colored curves are the model results.

The models were created using GEANT 3. As the positrons annihilate, they primarily emit two photons, in opposite directions, each with an energy of 511 keV, the rest mass energies of the electron and positron. The light blue curve (labeled 0 m) is for a positron source located immediately outside the aircraft. For this model, the 511 keV photons strike the aircraft directly without passing through any surrounding air. The simulations include the material in the aircraft and the instrument. In the simulations, some of the photons strike and deposit all of their energy in the NaI crystals, resulting in the line seen at 511 keV in the figure. The other curves are for a uniform and isotropic source of stationary positrons in the air surrounding the aircraft out to the specified radius. An isotropic source was chosen because a source located just above or below the aircraft would not result in nearly the same count rates in the top and bottom detectors, as was observed. Photons Compton scatter in the air (extended sources only), the aircraft and the instrument material surrounding the detectors, or deposit only a fraction of their energy in the NaI crystal, resulting in the low-energy Compton component seen below the 511 keV line.

As can be seen, the source models with no air and with a source radius of 450 m are not consistent with the data, producing too few counts at low energies. The models with large source volumes (900 m and 2000 m) agree better with the data, approximately matching both the line emission and the low-energy Compton component. As a check, the aircraft mass model was doubled, and even in this unrealistic case, the local source at the aircraft could be ruled out. These results show that the observed enhancement is consistent with being mostly from 511 keV emission from positrons and that the emission originated in a large volume of air surrounding or next to the aircraft. Specifically, the source radius appears to be greater than approximately 450 m and is most consistent with a radius of more than 1 km.

The small enhancement in counts at energies above the 511 keV line seen in figures 4 and 5 is also visible in figure 6. Since we did not attempt to model this component, a small discrepancy between the models and the data exists at these higher energies.

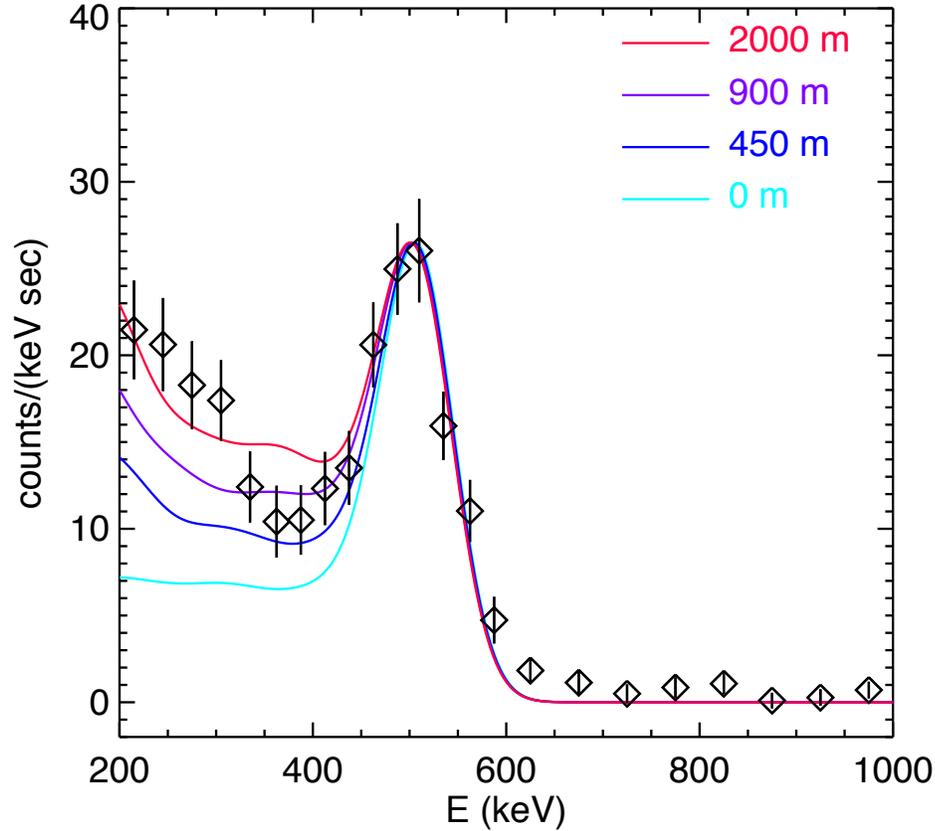

FIGURE 6. Background-subtracted energy spectrum showing the 511 keV enhancement. The data are the combined energy spectra of the two events, from both top and bottom NaI detectors, during times 0.8 – 0.9 sec in the records. The colored curves are the model results with the positrons filling a volume of air out to the specified radius in the figure.

The fit of the 900 m extended-source model (purple curve) found that for each event the GV was in or near a region with a positron annihilation rate of approximately $(2.2\pm0.7)\times10^{-4}$ /(cm³ sec). This corresponds to approximately $6.6\times10^{11}$ positron annihilations per second in the volume 900 m in radius, which is equivalent to approximately 18 Curies. The red curve corresponds to $(1.5\pm0.5)\times10^{-4}$ /(cm³ sec) and $5.0\times10^{12}$ positron annihilations per second in a 2000 m radius volume for each event, and the dark blue curve corresponds to $(5.8\pm2)\times10^{-4}$ /(cm³ sec) and $2.2\times10^{11}$ positron annihilations per second in a 450 m radius volume for each event.

## 5. Discussion

We are not aware of 511 keV enhancements within thunderstorms being reported before and so this appears to be a new phenomenon. Because the lifetime of positrons is very short and the emission for each event lasts approximately 0.2 seconds, there must be a source of positrons that lasts 0.2 seconds. Moreover, since the aircraft only moved tens of meters in the time that the 511 keV enhancements turned on and since the attenuation length of the 511 keV photons is ~500 m at that altitude, the change in the count rates could not be because the aircraft entered a spatial structure. Instead, the time profile must represent a temporal change in the density of positrons.

Positrons are commonly produced by pair production when gamma rays ($\varepsilon_\gamma > 2mc^2$) interact with atomic nuclei. These gamma rays may be provided by the bremsstrahlung emission of runaway electrons, as discussed above, or during cosmic-ray air showers. Alternatively, positrons may be emitted by radioactive decays. In any case, the positrons are usually created with relativistic energies above $\varepsilon_{th}$, and so the positrons may run away, similarly to the electrons but in the opposite direction. Even if the positrons are created with their momentum in the opposite direction of the electric field vector, a large fraction of the positrons may turn around and run away [Dwyer 2012]. It should be noted that the energy loss curve for positrons is very close to that of the electrons seen in figure 1, and so the physics of runaway positrons is similar to that of runaway electrons. One difference, however, is that there will be no avalanche multiplication of the positrons and so there is no RREA analogue for positrons. Because in a strong electric field the positrons quickly accelerate to highly relativistic energies (many tens of MeV) the annihilation cross-section is low and so they may propagate large distances, of the order of a kilometer through air at standard conditions and several kilometers at the altitude of the GV. The propagating beam of positrons will also generate bremsstrahlung gamma rays with energies reaching perhaps 100 MeV in thunderstorm electric fields.

We next discuss the three possible sources of the positrons.

*Positron Feedback*. Relativistic runaway electron avalanches create positrons that run away in the opposite direction, traveling hundreds of meters, if not kilometers. Indeed, this is the principle behind relativistic feedback. Runaway electrons produce gamma rays from bremsstrahlung interactions with air. Some of these gamma rays interact with atomic nuclei via pair production. Many of the positrons then turn around in the electric field and run away to the start of the avalanche region. After exiting the high-field region, the energetic positrons slow down and annihilate. Because many more bremsstrahlung gamma rays experience Compton scattering than pair production, a large flux of backscattered photons will also be present. These backscattered photons, which mostly have energies below a few hundred keV, may interact with air and produce additional seed runaway electrons. This is the principle behind gamma-ray feedback (also called x-ray feedback). ADELE did not record a large flux of low-energy photons during the 511 keV enhancements, which suggests that, if the 511 keV events were produced by relativistic feedback, ADELE must have been far from the end of the avalanche region where the bulk of the runaway electrons were produced. Furthermore, a nearby source of low-energy photons, such as backscattered photons from RREAs, would produce very different count rates on the top and bottom detectors in the 100- 400 keV range if the source region were either above or below the aircraft, which was not observed. In addition, as the positrons propagate, they generate high-energy bremsstrahlung photons. Because ADELE only recorded a small enhancement at high energies, it could not have been within the positron beam, since it would then have recorded these high-energy photons and the energetic positrons directly.

One possible scenario is that as ADELE approached a gamma-ray glow source region, the electric field experienced by the aircraft also increased. When the field reached a large enough value, an electrical discharge was initiated from the aircraft, moving electric charge and temporarily steering the runaway positrons closer to the aircraft. Based upon the altitude of the aircraft, we believe that ADELE was between the upper negative screening layer of the cloud and the main positive charge region, and so the runaway electron avalanches would have been directed downward and the positrons directed upward. For event 3, the aircraft may have subsequently entered the runaway electron avalanche region, seen as event 4. However, no enhancement was seen before or after event 1, so, if present, the glow must have been weaker than event 4, since ADELE can record glows even without entering their source region.

The advantage of this mechanism is that relativistic feedback can produce almost arbitrarily large fluxes of positrons. The challenge will be to explain how the positrons are steered towards the aircraft without hitting it and without producing large fluxes of high-energy photons. It is also not clear why the two events are so similar.

*Altering the cosmic ray background.* If the GV initiated a positive leader that traveled some distance, the aircraft and its immediate surroundings might charge negative, drawing in atmospheric cosmic-ray secondary positrons and repelling atmospheric cosmic-ray secondary electrons, increasing the local density of positrons. Perhaps the increase in the high-energy positrons is offset by the decrease in high-energy electrons (Carlson et al. 2008), resulting in little change in the counts at higher energies. A leader discharge may have been the response of the aircraft entering a high-electric-field region. We note that the 511 keV count rate increase is almost simultaneous with large signals occurring on the flat-plate antenna, possibly indicating that the aircraft was experiencing electrical discharges. Because the production of positrons by cosmic rays is roughly constant with time, in order to increase the number of positrons near the aircraft, the positrons must have been steered in from a relatively large volume. The attenuation length of 511 keV photons is approximately 500 m at the 14.1 km altitude, so the positrons would have to propagate inward over a length scale larger than this. For instance, to increase the positron count rate by a factor of 12, as observed, positrons would need to be efficiently propagated inward over a volume of more than 1 km in radius. These positrons would generate bremsstrahlung gamma rays, similarly to the positron feedback case. A challenge for this model will be to explain how a large enough electric field could be generated to bring cosmic-ray positrons in from such a large distance and why only relatively small enhancements are seen at higher energies. As with the positron feedback case, it is not obvious why the two events would be so similar. If this scenario is correct, it is interesting that an aircraft could influence such a large volume of the thunderstorm. It is also interesting that the GV did this twice, if not three times in a 35 second time period.

*Radioactive Decays.* Some radioactive isotopes found inside thunderstorms are a source of positrons. A beam of runaway electrons and the resulting gamma rays could create a localized region of enhanced radioactivity. For example, $^{13}$N and $^{15}$O are both positron emitters, with half-lives of 10 min and 2 min respectively. Along with neutrons, they are byproducts of photonuclear interactions of the energetic gamma rays with air. Simulations of the photonuclear interactions of TGF gamma rays with air have found that somewhere between $10^{12}$ and $10^{15}$ neutrons are produced (Babich and Roussel-Dupré 2007; Carlson et al. 2010). The radioactive products such as $^{13}$N and $^{15}$O have not been investigated, as far as we know, but similarly large numbers of these radioactive isotopes should also be produced. Because runaway electrons or the bremsstrahlung that they produce would cause a large flux increase at higher energies, which was not seen by ADELE, if the 511 keV emission was from these radioactive decays, then the RREAs must have occurred earlier, before the aircraft was in the vicinity. On the other hand, since the half-lives of these radioactive isotopes are much longer than the 0.2 second durations of the 511 keV enhancements, there would need to be some mechanism for bringing the positrons closer to the aircraft. It should be noted that the mobility of ions is too low for them to drift a significant distance during the 0.2 seconds of the events, and so the time structure of the event cannot be from the motion of the radioactive isotopes. One possibility is that an electric field change caused by a discharge from the aircraft resulted is these positrons being steered towards the aircraft. During the radioactive decays, the positrons are emitted with energies above 1 MeV, and so they could easily run away in fields above the RREA threshold, traveling large distances from the source. However, the problem with bremsstrahlung photons produced by the energetic positrons, discussed above, also occurs with this mechanism.

Alternatively, it is conceivable that the aircraft was already in a region filled with the radioactive isotopes and the positrons were being accelerated away by a strong electric field before the events. When the discharge occurred, it could have shorted out the electric field, causing the positrons to stop running away and to decay locally. If the 511 keV emissions are caused by radioactive byproducts of RREAs, then this offers a new way of studying TGFs, which are fairly rare, since a balloon or aircraft could measure the 511 keV signature of TGFs well after the TGF occurred.

# 6. Summary

In summary, the ADELE instrument on a Gulfstream V aircraft detected two, possibly three, enhancements in 511 keV emissions inside an active thunderstorm, indicating that the aircraft was briefly emerged in or near clouds of positrons. It is not clear where the excess positrons came from, but the enhancements were accompanied by electrical activity near the aircraft, suggesting that the electrical environment near the aircraft may have played a role in creating the enhancements or bringing them nearer to the aircraft. However, detailed modeling is needed before conclusions can be drawn about the likely sources of the positrons.

This work was supported in part by NSF awards AGS 1519236, 1160226, 0619941 and 0846609 and NASA award DPR S-15633-Y.